\documentstyle[11pt]{article}

\begin{document}

\pagestyle{plain}


\catcode`@=11
\def\un#1{\relax\ifmmode\@@underline#1\else
        $\@@underline{\hbox{#1}}$\relax\fi}
\catcode`@=12


\let\under=\b                   
\let\ced=\c                     
\let\du=\d                      
\let\um=\H                      
\let\sll=\l                     
\let\Sll=\L                     
\let\slo=\o                     
\let\Slo=\O                     
\let\tie=\t                     
\let\br=\u                      


\def\a{\alpha}
\def\b{\beta}
\def\c{\chi}
\def\d{\delta}
\def\e{\epsilon}
\def\f{\phi}
\def\g{\gamma}
\def\h{\eta}
\def\i{\iota}
\def\j{\psi}
\def\k{\kappa}
\def\l{\lambda}
\def\m{\mu}
\def\n{\nu}
\def\o{\omega}
\def\p{\pi}
\def\q{\theta}
\def\r{\rho}
\def\s{\sigma}
\def\t{\tau}
\def\u{\upsilon}
\def\x{\xi}
\def\z{\zeta}
\def\D{\Delta}
\def\F{\Phi}
\def\G{\Gamma}
\def\J{\Psi}
\def\L{\Lambda}
\def\O{\Omega}
\def\P{\Pi}
\def\Q{\Theta}
\def\S{\Sigma}
\def\U{\Upsilon}
\def\X{\Xi}

\def\ve{\varepsilon}
\def\vf{\varphi}
\def\vr{\varrho}
\def\vs{\varsigma}
\def\vq{\vartheta}


\def\ca{{\cal A}}
\def\cb{{\cal B}}
\def\cc{{\cal C}}
\def\cd{{\cal D}}
\def\ce{{\cal E}}
\def\cf{{\cal F}}
\def\cg{{\cal G}}
\def\ch{{\cal H}}
\def\ci{{\cal I}}
\def\cj{{\cal J}}
\def\ck{{\cal K}}
\def\cl{{\cal L}}
\def\cm{{\cal M}}
\def\cn{{\cal N}}
\def\co{{\cal O}}
\def\cp{{\cal P}}
\def\cq{{\cal Q}}
\def\car{{\cal R}}
\def\cs{{\cal S}}
\def\ct{{\cal T}}
\def\cu{{\cal U}}
\def\cv{{\cal V}}
\def\cw{{\cal W}}
\def\cx{{\cal X}}
\def\cy{{\cal Y}}
\def\cz{{\cal Z}}


\def\Sc#1{{\hbox{\sc #1}}}      
\def\Sf#1{{\hbox{\sf #1}}}      



\def\slpa{\slash{\pa}}                            
\def\slin{\SLLash{\in}}                                   
\def\bo{{\raise-.5ex\hbox{\large$\Box$}}}               
\def\cbo{\Sc [}                                         
\def\pa{\partial}                                       
\def\de{\nabla}                                         
\def\dell{\bigtriangledown}                             
\def\su{\sum}                                           
\def\pr{\prod}                                          
\def\iff{\leftrightarrow}                               
\def\conj{{\hbox{\large *}}}                            
\def\ltap{\raisebox{-.4ex}{\rlap{$\sim$}} \raisebox{.4ex}{$<$}}   
\def\gtap{\raisebox{-.4ex}{\rlap{$\sim$}} \raisebox{.4ex}{$>$}}   
\def\TH{{\raise.2ex\hbox{$\displaystyle \bigodot$}\mskip-4.7mu \llap H \;}}
\def\face{{\raise.2ex\hbox{$\displaystyle \bigodot$}\mskip-2.2mu \llap {$\ddot
        \smile$}}}                                      
\def\dg{\sp\dagger}                                   
\def\ddg{\sp\ddagger}                                   
\def\di{\int\!\!\!\int}                                 

\font\tenex=cmex10 scaled 1200


\def\sp#1{{}^{#1}}                              
\def\sb#1{{}_{#1}}                              
\def\oldsl#1{\rlap/#1}                          
\def\slash#1{\rlap{\hbox{$\mskip 1 mu /$}}#1}      
\def\Slash#1{\rlap{\hbox{$\mskip 3 mu /$}}#1}      
\def\SLash#1{\rlap{\hbox{$\mskip 4.5 mu /$}}#1}    
\def\SLLash#1{\rlap{\hbox{$\mskip 6 mu /$}}#1}      
\def\PMMM#1{\rlap{\hbox{$\mskip 2 mu | $}}#1}   %
\def\PMM#1{\rlap{\hbox{$\mskip 4 mu ~ \mid $}}#1}       %
\def\Tilde#1{\widetilde{#1}}                    
\def\Hat#1{\widehat{#1}}                        
\def\Bar#1{\overline{#1}}                       
\def\bra#1{\left\langle #1\right|}              
\def\ket#1{\left| #1\right\rangle}              
\def\VEV#1{\left\langle #1\right\rangle}        
\def\abs#1{\left| #1\right|}                    
\def\leftrightarrowfill{$\mathsurround=0pt \mathord\leftarrow \mkern-6mu
        \cleaders\hbox{$\mkern-2mu \mathord- \mkern-2mu$}\hfill
        \mkern-6mu \mathord\rightarrow$}
\def\dvec#1{\vbox{\ialign{##\crcr
        \leftrightarrowfill\crcr\noalign{\kern-1pt\nointerlineskip}
        $\hfil\displaystyle{#1}\hfil$\crcr}}}           
\def\dt#1{{\buildrel {\hbox{.}} \over {#1}}}     
\def\dtt#1{{\buildrel \bullet \over {#1}}}              
\def\der#1{{\pa \over \pa {#1}}}                
\def\fder#1{{\d \over \d {#1}}}                 


\def\frac#1#2{{\textstyle{#1\over\vphantom2\smash{\raise.20ex
        \hbox{$\scriptstyle{#2}$}}}}}                   
\def\ha{\frac12}                                        
\def\sfrac#1#2{{\vphantom1\smash{\lower.5ex\hbox{\small$#1$}}\over
        \vphantom1\smash{\raise.4ex\hbox{\small$#2$}}}} 
\def\bfrac#1#2{{\vphantom1\smash{\lower.5ex\hbox{$#1$}}\over
        \vphantom1\smash{\raise.3ex\hbox{$#2$}}}}       
\def\afrac#1#2{{\vphantom1\smash{\lower.5ex\hbox{$#1$}}\over#2}}    
\def\partder#1#2{{\partial #1\over\partial #2}}   
\def\parvar#1#2{{\d #1\over \d #2}}               
\def\secder#1#2#3{{\partial^2 #1\over\partial #2 \partial #3}}  
\def\on#1#2{\mathop{\null#2}\limits^{#1}}               
\def\bvec#1{\on\leftarrow{#1}}                  
\def\oover#1{\on\circ{#1}}                              

\def\[{\lfloor{\hskip 0.35pt}\!\!\!\lceil}
\def\]{\rfloor{\hskip 0.35pt}\!\!\!\rceil}
\def\Lag{{\cal L}}
\def\du#1#2{_{#1}{}^{#2}}
\def\ud#1#2{^{#1}{}_{#2}}
\def\dud#1#2#3{_{#1}{}^{#2}{}_{#3}}
\def\udu#1#2#3{^{#1}{}_{#2}{}^{#3}}
\def\calD{{\cal D}}
\def\calM{{\cal M}}

\def\szet{{${\scriptstyle \b}$}}
\def\ulA{{\un A}}
\def\ulM{{\underline M}}
\def\cdm{{\Sc D}_{--}}
\def\cdp{{\Sc D}_{++}}
\def\vTheta{\check\Theta}
\def\gg{{\hbox{\sc g}}}
\def\fracm#1#2{\hbox{\large{${\frac{{#1}}{{#2}}}$}}}
\def\half{{\fracm12}}
\def\ha{\half}
\def\tr{{\rm tr}}
\def\Tr{{\rm Tr}}
\def\itrema{$\ddot{\scriptstyle 1}$}
\def\ula{{\underline a}} \def\ulb{{\underline b}} \def\ulc{{\underline c}}
\def\uld{{\underline d}} \def\ule{{\underline e}} \def\ulf{{\underline f}}
\def\ulg{{\underline g}}
\def\items#1{\\ \item{[#1]}}
\def\ul{\underline}
\def\un{\underline}
\def\fracmm#1#2{{{#1}\over{#2}}}
\def\footnotew#1{\footnote{\hsize=6.5in {#1}}}
\def\low#1{{\raise -3pt\hbox{${\hskip 0.75pt}\!_{#1}$}}}

\def\Dot#1{\buildrel{_{_{\hskip 0.01in}\bullet}}\over{#1}}
\def\dt#1{\Dot{#1}}
\def\DDot#1{\buildrel{_{_{\hskip 0.01in}\bullet\bullet}}\over{#1}}
\def\ddt#1{\DDot{#1}}



\newskip\humongous \humongous=0pt plus 1000pt minus 1000pt
\def\caja{\mathsurround=0pt}
\def\eqalign#1{\,\vcenter{\openup2\jot \caja
        \ialign{\strut \hfil$\displaystyle{##}$&$
        \displaystyle{{}##}$\hfil\crcr#1\crcr}}\,}
\newif\ifdtup
\def\panorama{\global\dtuptrue \openup2\jot \caja
        \everycr{\noalign{\ifdtup \global\dtupfalse
        \vskip-\lineskiplimit \vskip\normallineskiplimit
        \else \penalty\interdisplaylinepenalty \fi}}}
\def\li#1{\panorama \tabskip=\humongous                         
        \halign to\displaywidth{\hfil$\displaystyle{##}$
        \tabskip=0pt&$\displaystyle{{}##}$\hfil
        \tabskip=\humongous&\llap{$##$}\tabskip=0pt
        \crcr#1\crcr}}
\def\eqalignnotwo#1{\panorama \tabskip=\humongous
        \halign to\displaywidth{\hfil$\displaystyle{##}$
        \tabskip=0pt&$\displaystyle{{}##}$
        \tabskip=0pt&$\displaystyle{{}##}$\hfil
        \tabskip=\humongous&\llap{$##$}\tabskip=0pt
        \crcr#1\crcr}}


\def\NPB{{\sf Nucl. Phys. }{\bf B}}
\def\PL{{\sf Phys. Lett. }}
\def\PRL{{\sf Phys. Rev. Lett. }}
\def\PRD{{\sf Phys. Rev. }{\bf D}}
\def\CQG{{\sf Class. Quantum Grav. }}
\def\JMP{{\sf J. Math. Phys. }}
\def\SJNP{{\sf Sov. J. Nucl. Phys. }}
\def\SPJ{{\sf Sov. Phys. J. }}
\def\JETPL{{\sf JETP Lett. }}
\def\TMP{{\sf Theor. Math. Phys. }}
\def\IJMPA{{\sf Int. J. Mod. Phys. }{\bf A}}
\def\MPL{{\sf Mod. Phys. Lett. }}
\def\CMP{{\sf Commun. Math. Phys. }}
\def\AP{{\sf Ann. Phys. }}
\def\PR{{\sf Phys. Rep. }}

\def\app#1#2#3{Acta Phys.~Pol.~{\bf B{#1}} (19{#2}) #3}
\def\pl#1#2#3{Phys.~Lett.~{\bf {#1}B} (19{#2}) #3}
\def\np#1#2#3{Nucl.~Phys.~{\bf B{#1}} (19{#2}) #3}
\def\prl#1#2#3{Phys.~Rev.~Lett.~{\bf #1} (19{#2}) #3}
\def\pr#1#2#3{Phys.~Rev.~{\bf D{#1}} (19{#2}) #3}
\def\cqg#1#2#3{Class.~and Quantum Grav.~{\bf {#1}} (19{#2}) #3}
\def\cmp#1#2#3{Commun.~Math.~Phys.~{\bf {#1}} (19{#2}) #3}
\def\jmp#1#2#3{J.~Math.~Phys.~{\bf {#1}} (19{#2}) #3}
\def\ap#1#2#3{Ann.~Phys.~(NY)~{\bf {#1}} (19{#2}) #3}
\def\prep#1#2#3{Phys.~Rep.~{\bf {#1}C} (19{#2}) #3}
\def\ptp#1#2#3{Progr.~Theor.~Phys.~{\bf {#1}} (19{#2}) #3}
\def\ijmp#1#2#3{Int.~J.~Mod.~Phys.~{\bf A{#1}} (19{#2}) #3}
\def\mpl#1#2#3{Mod.~Phys.~Lett.~{\bf A{#1}} (19{#2}) #3}
\def\nc#1#2#3{Nuovo Cim.~{\bf {#1}} (19{#2}) #3}
\def\ibid#1#2#3{{\it ibid.}~{\bf {#1}} (19{#2}) #3}
\def\sjnp#1#2#3{Sov.~J.~Nucl.~Phys.{\bf {#1}} (19{#2}) #3}
\def\tmp#1#2#3{Theor.~Math.~Phys.{\bf {#1}} (19{#2}) #3}

\topmargin=0in                          
\headheight=0in                         
\headsep=0in                    
\textheight=9in                         
\footskip=4ex           
\textwidth=6in                          
\hsize=6in                              
\parskip=\medskipamount                 
\lineskip=0pt                           
\abovedisplayskip=1em plus.3em minus.5em        
\belowdisplayskip=1em plus.3em minus.5em        
\abovedisplayshortskip=.5em plus.2em minus.4em  
\belowdisplayshortskip=.5em plus.2em minus.4em  
\def\baselinestretch{1.2}       
\thicklines                         


\def\sect#1{\bigskip\medskip \goodbreak \noindent{\bf {#1}} \nobreak \medskip}
\def\refs{\sect{References} \footnotesize \frenchspacing \parskip=0pt}
\def\Item{\par\hang\textindent}
\def\Itemitem{\par\indent \hangindent2\parindent \textindent}
\def\makelabel#1{\hfil #1}
\def\topic{\par\noindent \hangafter1 \hangindent20pt}
\def\Topic{\par\noindent \hangafter1 \hangindent60pt}

\newcommand{\be}{\begin{equation}}
\newcommand{\ee}{\end{equation}}
\newcommand{\bear}{\begin{eqnarray}}
\newcommand{\ear}{\end{eqnarray}}
\newcommand{\no}{\noindent}

\newcommand{\intl}{\int\limits}
\renewcommand{\d}{{\rm d}}
\newcommand{\eqpt}{.}
\newcommand{\eqcomma}{,}
\renewcommand\cd{\!\cdot\!}
\renewcommand\D{\nabla}
\renewcommand\={\,=\,}
\renewcommand\un{\underline}
\newcommand\ba{\Bar}
\newcommand\De{\Delta}
\newcommand\lr{\leftrightarrow}
\newcommand\ra{\rightarrow}
\newcommand\el{\ell}
\newcommand\hn{\hat{n}}
\newcommand\hX{\hat{X}}
\renewcommand\ve{\ \rule[-1mm]{.6mm}{4.5mm}\ }
\newcommand\sq{\square}
\newcommand\ax{\approx}
\newcommand\HC{{\underline C}}
\renewcommand\tr{{\rm tr\,}}
\newcommand\sa{{\sf a}}
\newcommand\sA{{\sf A}}
\newcommand\B{{\sf B}}
\newcommand\C{{\sf C}}
\newcommand\hC{{\Hat{\sf C}}}
\newcommand\sE{{\sf E}}
\newcommand\sF{{\sf F}}
\newcommand\sG{{\sf G}}
\newcommand\sI{{\sf I}}
\newcommand\R{{\sf R}}
\newcommand\hR{{\Hat{\sf R}}}
\newcommand\HR{{\Hat{    R}}}
\newcommand\sS{{\sf S}}
\newcommand\Ric{{\sf Ric}}
\renewcommand\X{{\sf X}}
\newcommand\Y{{\sf Y}}
\newcommand\Z{{\sf Z}}
\newcommand\og{{\,{\oover =}\,}}
\newcommand\da{\sp\dagger}
\newcommand\ov{\over}
\renewcommand\ch{\choose}

\renewcommand{\theequation}{\arabic{section}.\arabic{equation}}
\renewcommand{\arraystretch}{2.5}
\def\Eins{\mathord{1\hskip -1.5pt
\vrule width .5pt height 7.75pt depth -.2pt \hskip -1.2pt
\vrule width 2.5pt height .3pt depth -.05pt \hskip 1.5pt}}
\newcommand{\symb}{\mbox{symb}}
\renewcommand{\arraystretch}{2.5}
\def\non{\nonumber}
\def\beqn*{\begin{eqnarray*}}
\def\eqn*{\end{eqnarray*}}
\def\sy{\scriptscriptstyle}
\def\footstrut{\baselineskip 12pt}
\def\square{\kern1pt\vbox{\hrule height 1.2pt\hbox{\vrule width 1.2pt
   \hskip 3pt\vbox{\vskip 6pt}\hskip 3pt\vrule width 0.6pt}
   \hrule height 0.6pt}\kern1pt}
\def\half{{1\over 2}}
\def\third{{1\over3}}
\def\fourth{{1\over4}}
\def\fifth{{1\over5}}
\def\sixth{{1\over6}}
\def\seventh{{1\over7}}
\def\eigth{{1\over8}}
\def\ninth{{1\over9}}
\def\tenth{{1\over10}}
\def\ldop{\hbox{$\lbrace\mskip -4.5mu\mid$}}
\def\rdop{\hbox{$\mid\mskip -4.3mu\rbrace$}}
\def\arraystretch{2.5}
%
\def\bbbr{{\rm I\!R}}
\def\bbbone{{\mathchoice {\rm 1\mskip-4mu l} {\rm 1\mskip-4mu l}
{\rm 1\mskip-4.5mu l} {\rm 1\mskip-5mu l}}}
\def\bbbz{{\mathchoice {\hbox{$\sf\textstyle Z\kern-0.4em Z$}}
{\hbox{$\sf\textstyle Z\kern-0.4em Z$}}
{\hbox{$\sf\scriptstyle Z\kern-0.3em Z$}}
{\hbox{$\sf\scriptscriptstyle Z\kern-0.2em Z$}}}}
%

\pagestyle{empty}

\renewcommand{\thefootnote}{\fnsymbol{footnote}}

\hskip 9cm {\sl DESY 97-254 }
\vskip-.3cm
\hskip 9cm {\sl ANL-HEP-PR-97-99}
\vskip-.3cm
\hskip 9cm {\sl  MZ-TH/97-38} 

\vskip .8cm
\begin{center}
{\bf\Large A Closed Formula for the Riemann
Normal Coordinate Expansion}\\[1ex]
\vskip1.5cm

{\large Uwe M\"uller}
\\[1.5ex]
{\it 
Institut f\"ur Physik,
Johannes-Gutenberg-Universit\"at Mainz,\\
Staudingerweg 7,
D-55099 Mainz, Germany\\
e-mail: umueller@thep.Physik.Uni-Mainz.de
\medskip\\}
\vskip.3cm
{\large Christian Schubert
}\\[1.5ex]
{\it High Energy Physics Division, Argonne National Laboratory,\\
Argonne, IL 60439-4815, USA
\medskip\\}
and\\
{\it Institut f\"ur Physik,
Humboldt Universit\"at zu Berlin,\\
Invalidenstr.\ 110, D-10115 Berlin, Germany\\
e-mail: schubert@qft2.physik.hu-berlin.de
\medskip\\}
\vskip.3cm
{\large Anton E. M. van de Ven}
\\[1.5ex]
{\it II Institut f\"ur Theoretische Physik, Universit\"at Hamburg,\\
Luruper Chaussee 149, D-22761 Hamburg, Germany\\
e-mail: vandeven@mail.desy.de
\medskip\\}

\vskip1.2cm

{\large\bf Abstract}

\end{center}

\begin{quotation}

\noindent
We derive an integral representation which encodes
all coefficients of the Riemann normal coordinate
expansion and also a closed formula for those
coefficients.

\end{quotation}

\clearpage

\renewcommand{\thefootnote}{\protect\arabic{footnote}}
\pagestyle{plain}
\setcounter{page}{1}
\setcounter{footnote}{0}

\renewcommand{\theequation}{\arabic{equation}}
\setcounter{equation}{0}

In gauge theory, one 
often uses Fock-Schwinger gauge
~\cite{fock,schwinger}
to achieve manifest covariance 
in the calculation of effective actions,
anomaly densities, or other quantities. 
Fock-Schwinger gauge
``centered at $0$'' 
can be defined by the condition
\be
y^{\mu}A_{\mu}(y) = 0\eqpt
\label{definefockschwinger}
\ee\no
Locally in some neighbourhood of $0$, this condition 
can be solved in terms of
the following integral representation connecting the
gauge potential and the field strength tensor
~\cite{fascty,shifman},
\begin{equation}
A_{\mu}(y) = y^{\rho}
\int_0^1 \d v v F_{\rho\mu}(v y)
\label{AtoF}
\end{equation}
\no
(see our eqs. (9-13)).
More explicitly, in this gauge one can express the
coefficients of the 
Taylor expansion of $A$ at $x=0$
in terms of the covariant derivatives of $F$,
\begin{equation}
A_\mu(y)
=
\sum_{n=0}^{\infty}
{{(y\cdot \partial_x)}^n
\over n!(n+2)}
y^{\rho}
F_{\rho\mu}(0)
=
\sum_{n=0}^{\infty}
{{(y\cdot D_x)}^n
\over n!(n+2)}
y^{\rho}
F_{\rho\mu}(0)\eqpt
\label{Aexpand}
\end{equation}
Despite of the direct relation between
eqs.\ (\ref{AtoF}) and (\ref{Aexpand}),
the former version turns out to have advantages for
certain types of calculations in Yang-Mills
theory ~\cite{shifman,ss}. 

The analogue to Fock-Schwinger gauge in
gravity is the choice of a Riemann 
normal coordinate
system (see, e.\ g., ~\cite{eisenhart,alfrmu,luescher,friedan,frivdv}).
A normal coordinate system on a Riemannian manifold
centered at $0$ can be defined by 
\be
g_{\mu\nu}(0)=\delta_{\mu\nu},
\quad
y^{\mu}g_{\mu\nu}(y)=y^{\mu}g_{\mu\nu}(0)\eqpt
\label{def1riemann}
\ee\no
Alternatively, the second condition may be
replaced by the
following equivalent condition on the
affine connection
\be
y^{\mu}y^{\nu}
\Gamma_{\mu\nu}^{\lambda}(y)
=
0
\eqpt
\label{def2riemann}
\ee\no
Here
$\Gamma_{\mu\nu}^{\lambda}$ denotes
the Christoffel symbol for the Levi-Civita connection.
Locally this condition determines the coordinate system
up to a rigid rotation. 
The second condition
clearly shows the similarity to 
the Fock-Schwinger gauge
eq.\ (\ref{definefockschwinger}).
It also shows that in those coordinates
straight lines running through the origin
parametrize geodesics.

In Riemann normal coordinates the
Taylor coefficients of the metric tensor at $0$
can be expressed in terms of the covariant derivatives
of the Riemann curvature tensor. The Taylor
expansion starts as follows%
\footnote{
Our conventions are 
$$R^{\alpha}_{\,\,\,\beta\gamma\delta}
=
\Gamma^{\alpha}_{\,\,\,\beta\delta ,\gamma}
-
\Gamma^{\alpha}_{\,\,\,\beta\gamma ,\delta}
+
\Gamma^{\mu}_{\,\,\,\beta\delta}
\Gamma^{\alpha}_{\,\,\,\mu\gamma}
-
\Gamma^{\mu}_{\,\,\,\beta\gamma}
\Gamma^{\alpha}_{\,\,\,\mu\delta}
,
\quad
R_{\alpha\beta}=R^{\gamma}_{\,\,\,\alpha\gamma\beta}
,
\quad
R=R^{\mu}_{\,\,\,\mu}
.
\qquad\qquad
$$
},
\bear
g_{\mu\nu}(y)&=&\delta_{\mu\nu}
+\third R_{\mu\alpha\beta\nu}(0)y^{\alpha}y^{\beta}
+\sixth {\nabla}_{\gamma}R_{\mu\alpha\beta\nu}(0)
y^{\alpha}y^{\beta}
y^{\gamma}
\nonumber\\
&&
+{2\over 45}R_{\mu\alpha\beta\lambda}(0)
R^{\lambda}_{\,\,\,\gamma\delta\nu}(0)
y^{\alpha}y^{\beta}y^{\gamma}y^{\delta}
+{1\over 20}  
R_{\mu\alpha\beta\nu ;\gamma\delta} (0) y^\alpha y^\beta 
y^\gamma y^\delta 
+ O(y^5)
\label{rnclowest}
\ear\no
(in these coordinates
and at the origin there is no need to distinguish
between co- and contravariant indices). 

Riemann normal coordinates are a standard tool 
in the proof of differential geometric identities. 
To cite a prominent example,
they play a pivotal role in the local heat equation proof
of the Atiyah-Singer index theorem ~\cite{atbopa}.
In physics they are, for example, widely used for
nonlinear $\sigma$-model
calculations in curved spacetime backgrounds
~\cite{alfrmu,friedan,frivdv,basvan}.

\no
By differentiation of eq.\ (\ref{rnclowest}) one obtains a similar
expansion for $\Gamma_{\mu\nu}^{\lambda}$,
\be
\Gamma_{\mu\nu}^{\lambda}
(y)
=
{1\over 3}
\Bigl(R_{\nu\alpha\mu}^{\,\,\,\,\,\,\,\,\,\,\lambda}(0)
+
R_{\mu\alpha\nu}^{\,\,\,\,\,\,\,\,\,\,\lambda}(0)
\Bigr)
y^{\alpha}
+\ldots\eqpt
\label{Gammalowest}
\ee\no
In contrast to
its gauge theory analogue eq.\ (\ref{Aexpand}),
this expansion contains arbitrary powers of the
curvature tensor. As far as is known to the authors,
neither a closed formula for its coefficients has been
given in the literature, nor an equivalent of the
integral representation eq.\ (\ref{AtoF}). It is the purpose
of the present note to fill this gap.

In the literature one finds various ways of deriving
eq.\ (\ref{rnclowest})
from eqs.\ (\ref{def1riemann}) or (\ref{def2riemann}),
going back at least to 1925
~\cite{friedan,herglotz,oconnor,ambeoc}. We will
essentially
follow \cite{ambeoc} in the first part of our
argument.
First notice that
it suffices to find the Taylor expansion of the vielbein
$e^a_{\mu}(x)$, since 
\be
g_{\mu\nu}(x)=e_{\mu}^a(x)e_{\nu}^b(x)\delta_{ab}\eqpt
\label{g=esquared}
\ee\no
Finding an expression for the vielbein in terms of the
curvature effectively involves a twofold integration. As a first step,
we express the vielbein
connection $\omega$ in terms of the curvature.
For this purpose, consider the Lie transport of 
$\omega$ with respect to the radial vector
$y$. Writing the Lie derivative 
in terms of the interior product $i_y$ and the
exterior derivative $\d$,
\be
{\cal L}_y=i_y \d + \d i_y\eqcomma
\label{defLie}
\ee\no
one has
\be
{\cal L}_y\omega=i_y\d\omega+\d(i_y\omega)\eqpt
\label{Lomega}
\ee\no
We choose synchronous gauge,
i.e. Fock-Schwinger gauge for the vielbein
connection
\be
i_y\omega=0
\label{gaugecon}
\ee\no
This removes the
second term. Using the Cartan structure equation
$R=\d\omega +\omega\wedge\omega$ one obtains
\be
{\cal L}_y\omega^{\,\,\,a}_{\mu\,\,\,b}\d y^{\mu}
=
y^{\nu}R_{\,\,\,b\nu\mu}^{a}
\d y^{\mu}\eqpt
\label{Lomega=R}
\ee\no
A Taylor expansion of both sides of this
equation at $0$ then yields
\be
\omega^{\,\,\,a}_{\mu\,\,\,b}(y)
=\sum_{n=0}^{\infty}
{{(y\cdot\partial_x)}^n\over
n!(n+2)}
y^{\nu}
R_{\,\,\,b\nu\mu}^{a}(0)
=
\sum_{n=0}^{\infty}
{{(y\cdot\nabla_x)}^n\over
n!(n+2)}
y^{\nu}
R_{\,\,\,b\nu\mu}^{a}(0)
\eqpt
\label{omegathroughR}
\ee\no
This is, of course, nothing but the gauge theory
identity eq.\ (\ref{Aexpand}),
specialized to the $SO(D)$ case.

Next we act twice with ${\cal L}_y$ on $e$. Using the
absence of torsion, $\d e+\omega\wedge e=0$, as well as the
gauge condition eq.\ (\ref{gaugecon}), one finds
\bear
{\cal L}_ye
&=&
\omega
i_ye
+\d i_ye\eqcomma
\label{LE}\\
{\cal L}_y{\cal L}_ye
&=&
({\cal L}_y\omega)
i_ye
+
\omega
{\cal L}_y
i_ye
+
{\cal L}_y\d i_ye\eqpt
\label{LLE}
\ear\no
Using the gauge condition for the vielbein
\be
i_y e^a=\delta^a_{\mu}y^{\mu}\eqcomma
\label{gaugecond2}
\ee\no
eqs.\ (\ref{Lomega=R}), (\ref{LE}) and (\ref{LLE})
can be combined to yield
\be
({\cal L}_y-1){\cal L}_y e^a
=
R^{a}_{\,\,\,\mu\nu d}y^{\mu}y^{\nu}e^d\eqpt
\label{master}
\ee\no
On the left hand side 
one can trivially rewrite
\be
({\cal L}_y-1){\cal L}_y e_{\mu}\d y^{\mu}
=
\Bigl[(y\cdot\partial)(y\cdot \partial+1)
 e_{\mu}\Bigr]\d y^{\mu}\eqpt
\label{rewritemaster}
\ee\no
A Taylor expansion of both sides of eq.\ (\ref{master})
at $0$ then yields ~\cite{ambeoc}
\be
{(y\cdot \nabla)}^ke^a_{\mu}(0)
=
{k-1\over k+1}
{(y\cdot\nabla)}^{k-2}
\Bigl[R^{a}_{\,\,\,\alpha\beta b}y^{\alpha}y^{\beta}
e^b_{\mu}\Bigr](0)\eqcomma
\label{abc}
\ee\no
which expresses the Taylor coefficients of the vielbein
in terms of the covariant derivatives of the
Riemann tensor at $0$.

\noindent
Next we note that eq.\ (\ref{abc}) can be 
integrated to the following integral equation,
\be
e^a_{\mu}(y)
=
\delta^a_{\mu}+
y^{\alpha}y^{\beta}
\int_0^1
\d v v(1-v)
R^{a}_{\,\,\,\alpha\beta b}(vy)e^b_{\mu}(vy).
\label{intequ}
\ee\no
We decompose $e(y)$ as
\be
e(y)= \sum_{k=0}^{\infty}e_{(k)}(y)
\label{decomposeE}
\ee\no
where $k$ denotes the number of Riemann tensors
appearing in a given term in the normal coordinate
expansion of $e$. Obviously $e_{(k)}$ can be obtained
by iterating $k$ times eq.\ (\ref{intequ}),
and then replacing, under the integral, $e^b_{\mu}$ 
by $e^b_{\mu}(0)=\delta^b_{\mu}$.
This yields
\bear
e_{(k)}(y)&=&
\intl_{0}^{1}\d v_1(1-v_1)
\ldots
\intl_{0}^{1}\d v_k(1-v_k)v_1^{2k-1}v_2^{2k-3}\ldots v_k
\nonumber\\
&&\quad\times\,{\cal R}(v_1y,y)
{\cal R}(v_1v_2y,y)
\ldots{\cal R}(v_1v_2\ldots v_ky,y)\eqpt
\label{Vk}
\ear\no
Here we have introduced the abbreviation
\begin{equation}
{\cal R}^a{}_b(x,y)\equiv R^{a}_{\,\,\,\alpha\beta b}(x)
y^{\alpha}y^{\beta}\eqpt
\end{equation}
\no
To arrive at the metric itself, we need also the transposed of 
eq.\ (\ref{Vk}), which we can write as
\begin{eqnarray}
\hspace{-3ex}e_{(k)}^t(y)
\hspace{-1ex}&=&\hspace{-1ex}
\intl_{0}^{1}\d v_1(1-v_1)
\ldots
\intl_{0}^{1}\d v_k(1-v_k)v_1v_2^3\ldots v_k^{2k-1}
\nonumber\\
&&\quad\times\,{\cal R}(v_1v_2\ldots v_ky,y)
{\cal R}(v_2\ldots v_ky,y)
\ldots{\cal R}(v_ky,y)\eqpt
\end{eqnarray}
\no
The final result for the metric $g=e^te$ becomes
\begin{eqnarray}
\hspace{-3ex}g(y)\hspace{-1ex}&=&
\sum_{i=0}^{\infty}\sum_{j=0}^{\infty}
e^t_{(i)}(y)e_{(j)}(y)\nonumber\\
&=&
\sum_{k=0}^\infty
\intl_{0}^{1}\d v_1(1-v_1)
\intl_{0}^{1}\d v_2(1-v_2)
\ldots
\intl_{0}^{1}\d v_k(1-v_k)\nonumber\\
&&\quad\times\
\sum_{l=0}^k
v_1v_2^3\ldots v_l^{2l-1}
v_{l+1}^{2k-2l-1}v_{l+2}^{2k-2l-3}\ldots v_k\nonumber\\
&&\quad\times\,{\cal R}(v_1v_2\ldots v_ly,y)
{\cal R}(v_2\ldots v_ly,y)
\ldots{\cal R}(v_ly,y)\nonumber\\
&&\quad\times\,{\cal R}(v_{l+1}y,y)
{\cal R}(v_{l+1}v_{l+2}y,y)
\ldots{\cal R}(v_{l+1}\ldots v_ky,y)\eqpt
\label{g=intR}
\end{eqnarray}
\no
This integral representation, which for a given
value of $k$ encodes the coefficients of all terms in the
normal coordinate expansion having a fixed number of
Riemann tensors, appears to be the closest
possible analogue of the gauge theory formula eq.\ (\ref{AtoF})
\footnote
{After submitting this paper we were informed by Dolgov and Khriplovich
that they had published a similar result \cite{dolkri}.
However, their eq. (35)
concerns $x^\m\partial_\m g_{\a\b}$. A solution for the metric (or the
vielbein) itself is not given.}.

To explicitly obtain the coefficients, we now 
use eqs.(\ref{def2riemann}),(\ref{gaugecon})
again to covariantly
Taylor expand all Riemann tensor factors,
\begin{equation}
{\cal R}(vy,y)=\sum_{\kappa=0}^\infty\displaystyle
{\displaystyle v^\kappa(y\cdot\nabla)^\kappa\over{\displaystyle\kappa!}}
{\cal R}(0,y)\eqpt
\label{Rtaylor}
\end{equation}
\no
This leads to coefficient integrals which are easily
calculated, with the result
\begin{equation}
e_{(k)}(y)
=\sum_{\kappa_1,\ldots,\kappa_k=0}^\infty
{C_{\kappa_1,\ldots,\kappa_k}\over
{(\kappa_1+\ldots+\kappa_k+2k+1)!}}
{(y\cdot\nabla )}^{\kappa_1}{\cal R}(0,y)
\ldots{(y\cdot\nabla )}^{\kappa_k}{\cal R}(0,y)
\label{Vk-entw}
\end{equation}
\no
where
\begin{equation}\label{cformel}
C_{\kappa_1,\ldots,\kappa_k}=\prod_{l=1}^{k}
{\kappa_l+\kappa_{l+1}+\ldots+\kappa_k+2k-2l+1\choose\kappa_l}\eqpt
\end{equation}
\no
Introducing the Pochammer symbol $(a)_n=a(a+1)\ldots (a+n-1)$
this can also be written as
\be
e_{(k)}(y)
=\sum_{\kappa_1,\ldots,\kappa_k=0}^\infty
\prod_{l=1}^k
{{(y\cdot\nabla )}^{\kappa_l}{\cal R}(0,y)\over
{\kappa_l!(\kappa_l+\ldots+\kappa_k+2k-2l+2)_2}}\eqpt
\ee
\no
Neither the integral formula eq.\ (\ref{g=intR})
nor the coefficient formula eq.\ (\ref{cformel})
seem to have appeared in the literature before
(though ref.\ \cite{avramidi}
contains formulas equivalent to  eq.\ (\ref{cformel})).
In ref.\ \cite{ambeoc} eq.\ (\ref{abc}) was instead used to
derive a recursion formula for the 
normal coordinate expansion coefficients.
Define the matrices \cite{vandeven}
\be
\sE_k \= (e_{\mu}^a)_{,\m_1\dots\m_k}(0)\eqpt
\label{defEk}
\ee\no
These are the $k$-th partial derivatives of the vielbein evaluated at the 
origin of the normal coordinate system. Define also 
\be
\R_k=R^{\mu\,\,\,\,\,\,\,\,\,\,\,\,\nu}
_{\,\,(\mu_1\mu_2\,\,\,;\ldots\mu_k)}(0)
\label{defRk}
\ee\no
which one can consider as (symmetric) matrices
in the indices $\mu$ and $\nu$.
Then we can rewrite eq.\ (\ref{abc}) as
\be
(k+1)\sE_k \= (k-1)\R_k\, +\,\sum_{n=2}^{k-2}\ {k-1\ch n+1}\  
\R_{k-n}\ (n+1)\sE_n 
\label{recurs}
\ee\no
for $k\ge 2$, with

\be
E_0=\Eins,\quad E_1 =0\eqpt
\label{E01}
\ee\no
Here $\Eins$ denotes the $D$ - dimensional unit matrix, and
symmetrization on the $k$ indices is understood for each summand.
This recursion relation was used in ~\cite{ambeoc}
to list,
in their eq.\ (2.6), the normal coordinate
expansion through 8-th order
(with several errors at 8-th order).
Our eqs.\ (\ref{Vk-entw}),  (\ref{cformel})
resolve this recursion, as can be easily
verified by rewriting them in the following form,
\be
(k+1)\sE_k\= (k-1)\R_k\, +\sum_{\b=1}^\infty\Big[\prod_{\a=1}^\b
\sum_{n_\a=2}^{n_{\a-1}-2}\ {n_{\a-1}-1\ch n_\a+1}\ \R_{n_{\a-1}-n_\a}\Big] 
(n_\b -1)\R_{n_\b}
\label{solverecurs}
\ee\no
(with $n_0=k$).

\noindent
To tenth order the coefficients are given explicitly in
the appendix.

While the existence of a 
non-recursive formula for the normal coordinate
coefficients may be of independent mathematical
interest, we expect it also to
become of practical use in computer-based
high-order calculations of physically
interesting quantities. 
In particular, in recent years rapid progress has
been made in
the calculation of heat-kernel 
coefficients and effective actions.
Due to improvements
in computerization and to the availability of new
algorithms,
this type of calculation can now be extended to
orders which would have been unthinkable
a few years ago
~\cite{vandeven,fhss,mueller}.
This was also the original motivation for this work.
Of course the uses of the normal coordinate expansion
in physics are not restricted to quantum field theory;
for example,
our formulas may also be of relevance for the investigation
of the validity of Huygen's principle in curved spaces
(see, e.g., ref. [21] in \cite{vandeven}).

\vskip.5cm
\noindent
{\bf Acknowledgements:}
The authors would like to thank 
I.G. Avramidi, F. Bastianelli,
and R. Schimming for 
various communications.

\vfill\eject
{\bf Appendix: Coefficients to Tenth Order}

\renewcommand{\theequation}{A.\arabic{equation}}
\setcounter{equation}{0}
\noindent

To tenth order in the normal coordinate
expansion we find for the vielbein $e$ 
(writing $\{A\} = A + A^t$ and,
{\it par abus de language}, $\R_k$ for $\R_k y^k$) 
$$\eqalign{
e(y) &\= \sI +\frac1{3!}\R_2 +\frac1{4!} 2\R_3 
+\frac1{5!} (3\R_4 +\R_2\sp{2}) 
+\frac1{6!} (4\R_5 + 4\R_3\R_2 +2\R_2\R_3) \cr 
&
+\frac1{7!}(5\R_6 +10\R_4\R_2+10\R_3\sp{2}+\R_2(3\R_4+\R_2\sp{2}))\cr 
&
+\frac1{8!} (6\R_7 +20\R_5\R_2 +30\R_4\R_3 +6\R_3 (3\R_4 +\R_2\sp{2}) 
              +\R_2 (4\R_5 + 4\R_3\R_2 +2\R_2\R_3)) \cr 
&
+\frac1{9!} (7\R_8 +35\R_6\R_2 +70\R_5\R_3 +21\R_4(3\R_4 +\R_2\sp{2})
              + 7\R_3 (4\R_5 + 4\R_3\R_2 +2\R_2\R_3) \cr 
&\qquad
 +\R_2 (5\R_6 +10\R_4\R_2 +10\R_3\sp{2} +\R_2(3\R_4+\R_2\sp{2})) \cr 
&
+\frac1{10!} (8\R_9 +56\R_7\R_2 +140\R_6\R_3 +56\R_5(3\R_4+\R_2\sp{2})
       + 28\R_4(4\R_5+ 4\R_3\R_2 +2\R_2\R_3) \cr 
&\qquad
 +8\R_3(5\R_6 + 10\R_4\R_2 +10\R_3\sp{2} +\R_2(3\R_4+\R_2\sp{2})) \cr
&\qquad 
+\R_2 (6\R_7+20\R_5\R_2+30\R_4\R_3+6\R_3 (3\R_4+\R_2\sp{2}) 
    +\R_2 (4\R_5 + 4\R_3\R_2 +2\R_2\R_3)) \cr 
&
+\frac1{11!} (9\R_{10} +84\R_8\R_2 +252\R_7\R_3
               +126\R_6(3\R_4+\R_2\sp{2})  \cr 
&\qquad
+84\R_5 (4\R_5 + 4\R_3\R_2 +2\R_2\R_3)
+36\R_4(5\R_6 + 10\R_4\R_2 +10\R_3\sp{2} +\R_2(3\R_4+\R_2\sp{2})) \cr
&\qquad
+9  \R_3 (6\R_7 +20\R_5\R_2 +30\R_4\R_3 +6\R_3 (3\R_4 +\R_2\sp{2}) 
 +\R_2 (4\R_5 + 4\R_3\R_2 +2\R_2\R_3)) \cr 
&\qquad
+\R_2(7\R_8 +35\R_6\R_2 +70\R_5\R_3 +21\R_4(3\R_4 +\R_2\sp{2})
 + 7\R_3 (4\R_5 + 4\R_3\R_2 +2\R_2\R_3) \cr 
&\qquad
 +\R_2 (5\R_6 + 10\R_4\R_2 +10\R_3\sp{2} +\R_2(3\R_4+\R_2\sp{2}))) \cr
&
+\ O(11)
}$$
\no
and for the metric
$g = e^t e$
$$\eqalign{
g(y) &\= \sI + \frac1{2!}\,\frac23 \R_2 + \frac1{3!}\,\R_3 
+ \frac1{4!}\,\frac65 (\R_4 + \frac89\R_2\sp{2})
+ \frac1{5!}\,\frac43 (\R_5 + 2\{\R_2\R_3\}) \cr 
&
+ \frac1{6!}\,\frac{10}7(\R_6+\frac{17}5\{\R_2\R_4\}+\frac{11}2\R_3\sp{2} 
                      + \frac85\R_2\sp{3}) \cr 
&
+ \frac1{7!}\,\frac32 (\R_7 + \frac{46}9\{\R_2\R_5\} + 11\{\R_3\R_4\} 
+ \frac{62}9\R_2\R_3\R_2 +\frac{41}9\{\R_2\sp{2}\R_3\} ) \cr 
&
+ \frac1{8!}\,\frac{14}9 (\R_8 + \frac{50}7\{\R_2\R_6\} + 19\{\R_3\R_5\} 
         +\frac{126}5\R_4\sp{2} + \frac{130}7\R_2\R_4\R_2 \cr 
&\qquad
+\frac{339}{35}\{\R_2\sp{2}\R_4\} +14\R_3\R_2\R_3+\frac{163}7\{\R_2\R_3\R_3\} 
+\frac{128}{35}\R_2\sp{4} ) \cr 
&
+\frac1{9!}\,\frac85 (\R_9 +\frac{19}2\{\R_2\R_7\}+30\{\R_3\R_6\}
+49\{\R_4\R_5\}+40\R_2\R_5\R_2 +18\{\R_2\sp{2}\R_5\} + 85\R_3\sp{3}\cr &\qquad
+\frac{113}2\{\R_2\R_3\R_4\} +32\{\R_3\R_2\R_4\}+\frac{145}2\{\R_2\R_4\R_3\} 
+\frac{23}2\{\R_2\sp{3}\R_3\} +\frac{41}2\{\R_2\sp{2}\R_3\R_2\} )\cr
&
+\frac1{10!}\,\frac{18}{11} (\R_{10} +\frac{329}{27}\{\R_2\R_8\}
+\frac{89}2\{\R_3\R_7\} +86\{\R_4\R_6\} +\frac{952}9\R_5\sp{2}
+\frac{2030}{27}\R_2\R_6\R_2  \cr 
&\qquad
+\frac{829}{27}\{\R_2\sp{2}\R_6\}+ 78\R_4\R_2\R_4+\frac{593}3\{\R_4\R_4\R_2\}
+305\R_3\R_4\R_3 +\frac{443}2\{\R_3\sp{2}\R_4\} \cr 
&\qquad
+\frac{3164}{27}\{\R_2\R_3\R_5\}+\frac{575}9\{\R_3\R_2\R_5\}
+\frac{4775}{27}\{\R_2\R_5\R_3\}+\frac{245}9\{\R_2\sp{3}\R_4\} 
+\frac{1889}{27}\{\R_2\sp{2}\R_4\R_2\}\cr 
&\qquad
+\frac{4207}{54}\{\R_2\sp{2}\R_3\sp{2}\}
+\frac{1879}{27}\{\R_2\R_3\R_2\R_3\}
+\frac{3472}{27}(\R_2\R_3\sp{2}\R_2+\R_3\R_2\sp{2}\R_3)
+\frac{256}{27}\R_2^5 ) \cr
&
\ +\ O(11). 
}$$

\vfill\eject

\end{document}